\documentclass[11pt]{article}

\usepackage{geometry} 
\usepackage{amsmath} 
\usepackage{amsthm} 

\usepackage{amssymb} 
\usepackage{slashed}
\usepackage{dsfont}
\usepackage{bbold}
\usepackage[super]{nth}
\usepackage{blindtext}
\usepackage{graphicx}
\usepackage{bm}
\usepackage{color}   
\usepackage{hyperref}
\hypersetup{
    colorlinks=true, 
    linktoc=all,     
    linkcolor=blue,  
}
\usepackage{tensor} 

\author{Dominik Ciurla\thanks{domink.ciurla@student.uj.edu.pl}, Leszek Hadasz\thanks{leszek.hadasz@uj.edu.pl} \, and Thomas E. Williams\thanks{t.e.williams@doctoral.uj.edu.pl}  \\ Institute of Theoretical Physics, Jagiellonian University in Krak\'{o}w,\\ prof. {\L}ojasiewicza 11, 30-348 Kraków, Poland}

\title{\boldmath A Superspace Dirac Operator in NCG and the ``factorization" of the Ordinary Dirac Operator}
\begin{document}
\date{}

\maketitle

\abstract{We review a procedure of factorizing the Minkowski space Dirac operator over a~suitable superspace, discuss its Euclidean space version and apply
the worked out formalism in the case od an almost-commutative Dirac operator. The presented framework is an attempt to reconcile non-commutative geometry and supersymmetry.}
\flushbottom

\setcounter{tocdepth}{2}
\tableofcontents

\section{Introduction}

In 1928 P.A.M. Dirac reported his now famous procedure for deriving an equation governing the quantum mechanical properties for particles with half-integer spin \cite{dirac}.  The process he pioneered may be essentially described as taking the ``square root" of the Klein-Gordon equation.

The natural question, whether this process is iterable, was posed and solved by use of superspace coordinates and their (first order) derivatives  \cite{szwed}.  A series of papers followed, studying the free and interacting forms of the resulting equations acting on (super)spaces of superfields \cite{kp, kp2, bh, bs, szwed2}.

Recently, in \cite{williams} a procedure was proposed for the construction of physical models which exhibit supersymmetry within the framework of noncommutative geometry \cite{connes}.  Noncommutative geometry, pioneered by Alain Connes in the 1980s and 1990s, is a profoundly deep branch of mathematics with roots and branches stretching in many directions. The mathematical depth of this subject belies it's applicability to theoretical physics, where it has found great success in reproducing the Standard Model of particle physics coupled with gravity.  Specifically, it is within a certain subclass of noncommutative geometries known as almost commutative (AC) geometries, in which such physical models may be described.  This adaptation was pioneered in \cite{cl}, but for the working physicist we also recommend the presentation in \cite{wvs}.  Of central importance to this framework is the notion of a suitable Dirac operator.  Given that the natural geometric setting for supersymmetry is superspace \cite{martin, ggrs}, we expect that any Dirac operator which is claimed to govern the dynamics of particles in a supersymmetric model, should, in an essential way, take into account superspace coordinates and their derivatives.

One possibility would be to construct a superspace Dirac operator associated with the underlying superspace spin bundle.  This would be a sort of ``inside-out" approach where the fundamental space under consideration is a superspace exhibiting supersymmetry through infinitesimal global translations of its coordinates.  Considered in this way, supersymmetry is an explicit, unavoidable property of the model.  We postpone further discussion of this interpretation for future work.

Alternatively, inspired by the procedure outlined in \cite{szwed}, one may consider an ``outside-in" approach. This time, the basic ingredients are those of the usual AC-geometry approach for obtaining physical models from NCG, i.e. the underlying space is an ordinary Riemannian spin manifold and the Dirac operator is the spin connection acting fiberwise on square integrable sections of the spin bundle.  Supersymmetry and the gauge fields then emerge when considering the action of the ``square root" of the (unfluctuated!) total space Dirac operator on a restricted space of superfield spinors.

We now proceed to construct such an operator via this ``outside-in" approach.

\section{Factorization of the Dirac operator}
\subsection{Minkowski space -- the Szwed approach}
Using two-component spinor notation (ofttimes referred to as Van der Waerden notation) and the chiral representation for the Dirac matrices (for the conventions see \cite{WessAndBagger}),
one can write the Dirac equation in four dimensional Minkowski space  as
\begin{equation}
\label{Dirac}
-\left(
\begin{array}{cc}
i\bar\sigma^{\mu\,\dot\alpha\beta}\partial_\mu & m\delta^{\dot\alpha}_{\dot\beta} \\
m\delta^\beta_\alpha & i\sigma^\mu_{\alpha\dot\beta}\partial_\mu
\end{array}
\right)
\left(
\begin{array}{c}
\psi_\beta \\ \bar\chi^{\dot\beta}
\end{array}
\right)
\; \equiv \;
{\cal D}\left(
\begin{array}{c}
\psi\\ \bar\chi
\end{array}
\right)
\; = \; 0.
\end{equation}
Taking a ``square root'' of the Dirac operator corresponds to the construction of an operator, $A$, which satisfies
\begin{equation}
\label{prop1}
A^\dag\!A = {\cal D}.
\end{equation}
If one requires $A$ to be a local operator and to contain space-time derivatives, then, since there is no second order derivative in the Dirac operator, one is compelled to assume that the
coefficients of $\partial_\mu$ in $A$ are nilpotent.
Therefore one is lead to consider the operator $A$ as acting on a superspace
with the coordinates $(x^\mu, \theta^\alpha,\bar\theta^{\dot\alpha}).$

There are several first order differential operators which can be defined on this space. In particular, the spinorial ones,
\begin{eqnarray}
D_\alpha & = & \;\;\partial/{\partial\theta^\alpha} +
i\sigma^\mu{}_{\alpha\dot\alpha}\bar\theta^{\dot\alpha}\partial_\mu,
\nonumber \\
\bar D_{\dot\alpha} & = & \!-{\partial}/{\partial\bar\theta^{\dot\alpha}}
-i\theta^{\alpha}\sigma^\mu{}_{\alpha\dot\alpha}\partial_\mu.
\end{eqnarray}
satisfy an algebra with relations given by
\begin{eqnarray}
\label{algebra}
\left\{D_\alpha,D_\beta\right\}
& = &
\left\{\bar D_{\dot\alpha},\bar D_{\dot\beta}\right\}
= 0,
\nonumber \\
\left\{D_\alpha,\bar D_{\dot\beta}\right\}
& = & -2i\sigma^\mu{}_{\alpha\dot\beta}\partial_\mu.
\end{eqnarray}
If we now define $2\times 2$ matrices
\begin{equation}
\label{A:solution}
A_{\beta\alpha}
=
\left(
\begin{array}{rr}
D^\beta & -\bar D_{\dot\beta} \\
\bar D^{\dot\alpha} & D_\alpha
\end{array}
\right),
\end{equation}
then
\begin{equation}
(A_{\alpha\beta})^\dag A_{\beta\alpha}
=
\left(
\begin{array}{cc}
\{D^\beta,\bar D^{\dot\alpha}\} & \bar D_{\dot\beta} \bar D^{\dot\alpha} + D^\beta D_\alpha
\\[2pt]
\bar D_{\dot\beta} \bar D^{\dot\alpha} + D^\beta D_\alpha & \{D_\alpha,\bar D_{\dot\beta}\}
\end{array}
\right).
\end{equation}
In particular
\begin{equation}
\label{A:explicit}
(A_{\alpha\alpha})^\dag\!A_{\alpha\alpha}
=
-2\left(
\begin{array}{cc}
i\bar\sigma^{\mu\dot\alpha\alpha}\partial_\mu & M \\
M & i\sigma^\mu{}_{\alpha\dot\alpha}\partial_\mu
\end{array}
\right)
\end{equation}
with
\begin{equation}
\label{M:definition}
M \; = \; -\frac14\left(D\!D +\bar D\!\bar D\right) \equiv -\frac14\left(\bar D_{\dot\alpha} \bar D^{\dot\alpha} + D^\alpha D_\alpha\right).
\end{equation}

The equality (\ref{A:explicit}) was the motivation in \cite{szwed,bh} for postulating the following set of equations as a ``square root'' of the Dirac equation:
\begin{equation}
\label{Minkowski:basic:equations}
D^\alpha\psi_\alpha - \bar D_{\dot\alpha}\bar\chi^{\dot\alpha} = 0, \hskip 1cm \bar D^{\dot\alpha}\psi_\alpha +D_{\alpha}\bar\chi^{\dot\alpha} = 0,
\end{equation}
in which the spinors $\psi_\alpha$ and $\bar\chi^{\dot\alpha}$ are considered to be functions of the superspace coordinates $(x^\mu, \theta^\alpha,\bar\theta^{\dot\alpha}),$
and are subject to the additional constraint
\begin{equation}
\label{additional}
\left(D\!D +\bar D\!\bar D\right)\psi_\alpha + 4m\psi_\alpha =  \left(D\!D +\bar D\!\bar D\right)\bar\chi^{\dot\alpha} + 4m\bar\chi^{\dot\alpha} = 0.
\end{equation}

The solution set of these equations turned out to be nonempty and interesting. In particular, a simple case in which $\psi_\alpha = \chi_\alpha$
corresponds to the Maxwell superfield \cite{bh}.

\subsection{4d Euclidean space}
It is essential to the noncommutative methods, which we intend to employ in section 3, that the ``total-space" Dirac operator is Hermitian. Therefore we proceed in a Riemannian signature and for simplicity choose to work in 4-dimensional Euclidean space.

In particular, in this setting the Lorentz transformations are the 4-dimensional rotations characterized by the symmetry group $SO(4)$. Their spin representation is given by the universal covering Lie group, $\text{Spin(4)} \cong  SU(2)\times SU(2)$ and the corresponding Clifford algebra is isomorphic to the Lie algebra of infinitesimal generators, $\mathfrak{su}(2) \oplus \mathfrak{su}(2)$. \\After defining
\begin{equation}
\sigma^m \equiv (i\tau_1, i\tau_2, i\tau_3, \mathbf{1}_2) \quad \text{and} \quad \tilde\sigma^m \equiv (-i\tau_1, -i\tau_2, -i\tau_3, \mathbf{1}_2),
\end{equation}
where $\tau_i$ are the Pauli matrices, it is immediate to check that the Hermitian matrices
\begin{equation}
\gamma_{\rm E}^m \equiv \begin{pmatrix}
0&\sigma^m \\
\tilde\sigma^m&0
\end{pmatrix}
\end{equation}
generate the Clifford algebra of 4-dimensional Euclidean space,
\begin{equation}
\{ \gamma_{\rm  E}^m, \gamma_{\rm  E}^n \} = 2\delta^{mn}\mathbf{1}_4.
\end{equation}
Furthermore, this algebra possesses a natural grading induced by the operator
\begin{equation}
\label{Euclidean_gamma_5}
\gamma_{\rm E}^5 \equiv \gamma^1\gamma^2\gamma^3\gamma^4=\begin{pmatrix}
-\mathbf{1}_2&0\\
0&\mathbf{1}_2
\end{pmatrix}.
\end{equation}

The Euclidean Dirac operator has the form
\begin{equation}
\label{dfn:Dirac_operator_Euclidean}
{\cal D} = i \gamma_E^m \partial_m + m  \mathbf{1}_4= \left( \begin{matrix} m  \mathbf{1}_2 &i\sigma^m \partial_m \\ i \tilde{\sigma}^m \partial_m & m  \mathbf{1}_2 \end{matrix} \right)
\end{equation}
and acts on a bispinor
\begin{equation}
\Psi = \left( \begin{matrix} \psi \\ \tilde{\chi} \end{matrix} \right).
\end{equation}
As for the spinorial indices, we declare
\begin{equation}
\begin{split}
\psi = \left( \psi_\alpha \right) ,& \quad
\tilde{\chi} = \left( \tilde{\chi}^{\dot{\alpha}} \right), \\
\tilde{\sigma}^m = \left( \tilde{\sigma}^{m \dot{\alpha} \alpha} \right)&, \quad
\sigma^m = \left( \sigma^m_{\alpha \dot{\alpha}} \right)
\end{split}
\end{equation}
which allows us to present the Dirac equation as
\begin{equation}
\label{eq:Dirac_equation_Euclidean}
\begin{split}
i\tilde{\sigma}^{m \dot{\alpha} \alpha} \partial_m \psi_\alpha + m \tilde{\chi}^{\dot{\alpha}} &= 0,
\\[4pt]
i\sigma^m_{\alpha \dot{\alpha}} \partial_m \tilde{\chi}^{\dot{\alpha}} + m \psi_\alpha &= 0. 
\end{split}
\end{equation}

Unlike the Minkowski case, the spinors $\psi$ and $\tilde\chi$ transform independently under the action of $\text{Spin(4)}$. Indeed, if we parameterize
a matrix $L \in \text{SO}(4)$ as $L = \exp\omega$ (with $\omega_{mn} = - \omega_{nm}$) then
\begin{equation}
\psi'_\alpha(x) = M_\alpha^{\hskip 5pt\beta}\psi_\beta\big(L^{-1}x\big),
\hskip 1cm
\tilde\chi'{}^{\dot\alpha} = W^{\dot\alpha}_{\hskip 5pt\dot\beta} \tilde\chi^{\dot\beta}\big(L^{-1}x\big)
\end{equation}
where
\begin{equation}
\label{M:and:W}
M(L) = \exp \left( \textstyle{\frac{1}{8} }\omega_{mn}(\sigma^m \tilde{\sigma}^n - \sigma^n \tilde{\sigma}^m)\right), \quad
W(L) = \exp \left( \textstyle{\frac{1}{8}} \omega_{mn}(\tilde{\sigma}^m \sigma^n - \tilde{\sigma}^n \sigma^m) \right),
\end{equation}
are distinct operators.
i.e.\ $M(L)$ depends on $\omega_{mn}$ only through a combination
\begin{equation}
\sum\limits_{j=1}^3\Big(\sum\limits_{k=1}^3\sum\limits_{l=1}^{k-1}\epsilon_{jkl}\omega_{kl} + \omega_{j4}\Big)\tau_j
\end{equation}
while $W(L)$ depends on $\omega_{mn}$ through a combination
\begin{equation}
\sum\limits_{j=1}^3\Big(\sum\limits_{k=1}^3\sum\limits_{l=1}^{k-1}\epsilon_{jkl}\omega_{kl} - \omega_{j4}\Big)\tau_j.
\end{equation}

In order to construct a relevant superspace, we introduce two constant (anticommuting) spinors $\xi_\alpha$ and $\tilde\zeta^{\dot\alpha}.$  By construction,
under the action of $\text{Spin(4)}$ we see that
\begin{equation}
\xi_\alpha \to M_\alpha^{\hskip 5pt\beta}\xi_\beta,
\hskip 1cm
\tilde\zeta^{\dot\alpha} \to W^{\dot\alpha}_{\hskip 5pt\dot\beta}\tilde\zeta^{\dot\beta}
\end{equation}
and thus $\xi_\alpha$ and $\tilde\zeta^{\dot\alpha}$ are necessarily complex, i.e.\ we may treat $\xi_\alpha$ and
\(
\overline{\xi}^\beta = \left(\xi_\beta\right)^\dag,
\)
as well as $\tilde\zeta^{\dot\alpha}$ and
\(
\overline{\tilde\zeta}_{\dot\beta} = \left(\tilde\zeta^{\dot\beta}\right)^\dag,
\)
as independent Grasmann variables.

For the  Levi-Civita tensor we adapt the convention $\varepsilon_{12} = \varepsilon_{\dot1\dot2} = \varepsilon^{21}=\varepsilon^{\dot2\dot1}=1$. In effect
\[
\epsilon^{\alpha\beta}\epsilon_{\beta\gamma} = \delta^\alpha_\gamma,
\hskip 1cm
\epsilon_{\dot\alpha\dot\beta}\epsilon^{\dot\beta\dot\gamma} = \delta_{\dot\alpha}^{\dot\gamma}
\]
and
\begin{equation}
\varepsilon^{\dot\alpha\dot\beta}\varepsilon^{\alpha\beta}\sigma^m_{\beta\dot\beta}=\tilde\sigma^{m\dot\alpha\alpha}.
\end{equation}
Let us now define the spinorial derivatives
\begin{equation}
\label{spinorial:derivative:1a}
D^\alpha
=
\frac{\partial}{\partial\xi_\alpha} + i\,\overline{\tilde\zeta}_{\dot\beta}\,\tilde\sigma^{m\dot\beta\alpha}\,\partial_m,
\hskip 1cm
{\tilde D}_{\dot\alpha}
=
\frac{\partial}{\partial{\tilde\zeta}^{\dot\alpha}} + i\,{\bar\xi}^\beta\,\sigma^{m}_{\beta\dot\alpha}\,\partial_m,
\end{equation}
and consequently
\begin{equation}
\label{spinorial:derivative:1b}
{\overline D}_\alpha
=
\frac{\partial}{\partial{\bar \xi}^\alpha} + i\,\sigma^{m}_{\alpha\dot\beta}\,{\tilde\zeta}^{\dot\beta}\,\partial_m,
\hskip 1cm
{\overline{\tilde D}}{}^{\dot\alpha}
=
\frac{\partial}{\partial\overline{\tilde\zeta}_{\dot\alpha}} + i\,\tilde\sigma^{m\dot\alpha\beta}\,\xi_\beta\,\partial_m.
\end{equation}
They satisfy an algebra
\begin{equation}
\label{anticommutators}
\left\{D^\alpha,{\overline{\tilde D}}{}^{\dot\alpha}\right\} = 2i\, \tilde\sigma^{m\dot\alpha\alpha}\,\partial_m,
\hskip 1cm
\left\{{\overline D}_\alpha, {\tilde D}_{\dot\alpha}\right\} = 2i\,\sigma^{m}_{\alpha\dot\alpha}\,\partial_m,
\end{equation}
with all the remaining anticommutators vanishing. Moreover, if we  define
\begin{equation}
\label{supercharges:1a}
Q^\alpha
=
\frac{\partial}{\partial\xi_\alpha} - i\,\overline{\tilde\zeta}_{\dot\beta}\,\tilde\sigma^{m\dot\beta\alpha}\,\partial_m,
\hskip 1cm
{\tilde Q}_{\dot\alpha}
=
\frac{\partial}{\partial{\tilde\zeta}^{\dot\alpha}} - i\,{\bar\xi}^\beta\,\sigma^{m}_{\beta\dot\alpha}\,\partial_m,
\end{equation}
and
\begin{equation}
\label{supercharges:1b}
{\overline Q}_\alpha
=
\frac{\partial}{\partial{\bar \xi}^\alpha} - i\,\sigma^{m}_{\alpha\dot\beta}\,{\tilde\zeta}^{\dot\beta}\,\partial_m,
\hskip 1cm
{\overline{\tilde Q}}{}^{\dot\alpha}
=
\frac{\partial}{\partial\overline{\tilde\zeta}_{\dot\alpha}} - i\,\tilde\sigma^{m\dot\alpha\beta}\,\xi_\beta\,\partial_m,
\end{equation}
then it is immediate to check that all of the anticommutators involving one of the operators (\ref{spinorial:derivative:1a}) or (\ref{spinorial:derivative:1b}),
and one of the operators (\ref{supercharges:1a}) or (\ref{supercharges:1b}), vanish. In effect, all equations formulated in terms of derivatives
(\ref{spinorial:derivative:1a}) and (\ref{spinorial:derivative:1b}) are invariant under the (supersymmetry) transformations generated by
(\ref{supercharges:1a}) and (\ref{supercharges:1b}).

We next promote $\psi_\alpha$ and $\tilde\chi^{\dot\alpha}$ to spinor valued functions on the Euclidian superspace with coordinates
$(x,\xi_\alpha,\bar\xi^\alpha,\tilde\zeta^{\dot\alpha},\overline{\tilde\zeta}_{\dot\alpha})$ and, guided by (\ref{Minkowski:basic:equations}),
subject them to the following set of equations:
\begin{equation}
\label{euclidian:basic:equation:1}
D^\alpha\psi_{\alpha} + {\tilde D}_{\dot\alpha}\tilde\chi^{\dot\alpha} = 0
\end{equation}
and
\begin{equation}
\label{euclidian:basic:equation:2}
\overline{{\tilde D}}{}^{\dot\alpha}\psi_{\alpha} + \overline{D}_{\alpha}\tilde\chi^{\dot\alpha} = 0.
\end{equation}
In (\ref{euclidian:basic:equation:1}) the indices are summed over (so that the l.h.s.\ is a scalar) while (\ref{euclidian:basic:equation:2}) is a vanishing condition
for a certain tensor, and thus also has an invariant meaning.

From (\ref{anticommutators}), (\ref{euclidian:basic:equation:1}) and (\ref{euclidian:basic:equation:2}) we get
\begin{align}
i\tilde{\sigma}^{m \dot{\alpha} \alpha} \partial_m \psi_\alpha + \tilde M^{\dot\alpha}_{\hskip 5pt\dot\beta}\tilde{\chi}^{\dot{\beta}} &= 0, \label{Dirac_equation_Euclidean:2a}
\\[4pt]
i\sigma^m_{\alpha \dot{\alpha}} \partial_m \tilde{\chi}^{\dot{\alpha}} + M_\alpha^{\hskip 5pt\beta} \psi_\beta &= 0. \label{Dirac_equation_Euclidean:2b}
\end{align}
where
\begin{equation}
M_\alpha^{\hskip 5pt\beta} = \frac12\left(\delta_\alpha^\beta \,{\tilde D}_{\dot\alpha}\overline{{\tilde D}}{}^{\dot\alpha}+ \overline{D}_\alpha D^\beta\right),
\hskip 5mm
\tilde M^{\dot\alpha}_{\hskip 5pt\dot\beta} = \frac12\left(\delta^{\dot\alpha}_{\dot\beta}\,D^\alpha\overline{D}_\alpha + \overline{{\tilde D}}{}^{\dot\alpha}{\tilde D}_{\dot\beta}\right).
\end{equation}
We conclude that (\ref{euclidian:basic:equation:1}) and (\ref{euclidian:basic:equation:2}) imply Dirac equations for the (super) spinors $\psi_\alpha $ and $\tilde{\chi}^{\dot{\alpha}}$
on the subspace of superfields satisfying
\begin{equation}
\label{superfields:constraints}
M_\alpha^{\hskip 5pt\beta}\psi_\beta = m\psi_\alpha, \hskip 1cm \tilde M^{\dot\alpha}_{\hskip 5pt\dot\beta} \tilde{\chi}^{\dot{\beta}} = m \tilde{\chi}^{\dot{\alpha}}.
\end{equation}

To see that there exist nontrivial solutions of the set of equations (\ref{Dirac_equation_Euclidean:2a}), (\ref{Dirac_equation_Euclidean:2b}) and (\ref{superfields:constraints}) we consider
a simple case
\begin{equation}
\tilde{\chi}^{\dot{\alpha}} = \frac{\partial\psi_\alpha}{\partial\bar\xi^\beta} = \frac{\partial\psi_\alpha}{\partial\tilde\zeta^{\dot\beta}}= 0.
\end{equation}
Equation
\begin{equation}
\overline{{\tilde D}}{}^{\dot\alpha}\psi_{\alpha} = 0
\end{equation}
then implies that $\psi_\alpha$ depends on $\overline{\tilde\zeta}_{\dot\alpha}$ only through a combination of the form
\begin{equation}
y ^m = x^m - i\overline{\tilde\zeta}_{\dot\alpha}\tilde\sigma^{m\dot\alpha\alpha}\xi_\alpha.
\end{equation}
If we take
\begin{equation}
\label{an:Ansatz}
\psi_\alpha(x,\xi_\alpha,\tilde\zeta^{\dot\alpha}) = \lambda_\alpha (y) + F_{mn}(y)(\sigma^m\tilde\sigma^n)_\alpha^{\hskip 5pt\beta}\xi_\beta,
\end{equation}
then, since
\begin{equation}
D^\alpha y^m = 2i\overline{\tilde\zeta}_{\dot\alpha}\tilde\sigma^{m\dot\alpha\alpha}
\end{equation}
we get
\begin{equation}
\label{covariant:deriv:vanishing}
D^{\alpha}\psi_\alpha
=
\mathrm{Tr}(\sigma^m\tilde\sigma^n)\,F_{mn}(y)
+
2i\overline{\tilde\zeta}_{\dot\alpha}\tilde\sigma^{m\dot\alpha\alpha}\partial_m\lambda_\alpha (y)
+
2i\overline{\tilde\zeta}_{\dot\alpha}\xi_\beta (\tilde\sigma^p\sigma^m\tilde\sigma^n)^{\dot\alpha\beta}\partial_p F_{mn}(y).
\end{equation}
Vanishing of the second term on the r.h.s.\ of formula (\ref{covariant:deriv:vanishing}) implies that $\lambda_\alpha$ satisfies the massless Dirac equation,
\begin{equation}
i\tilde\sigma^{m\dot\alpha\alpha}\partial_m\lambda_\alpha = 0,
\end{equation}
meanwhile, vanishing of the first term implies that the tensor $F_{mn}$ is antisymmetric, and consequently the identity
\begin{equation}
\tilde\sigma^p\sigma^m\tilde\sigma^n = \epsilon^{pmnr}\,\tilde\sigma^r + \delta^{mp}\tilde\sigma^n + \delta^{mn}\tilde\sigma^p - \delta^{np}\tilde\sigma^m, \hskip 1cm \epsilon^{1234} = 1,
\end{equation}
applied to the last term, gives
\begin{equation}
\epsilon^{rpmn}\partial_pF_{mn} = 0, \hskip 1cm \partial^mF_{mn}=0.
\end{equation}
We conclude that a particular solution of the postulated set of equations is a spinor superfield with component fields consisting of a massless spinor field and a Maxwell gauge field.

Since the matrices (\ref{M:and:W}) are unitary with unit determinant, spinors $\xi^\alpha \equiv \epsilon^{\alpha\beta}\xi_\beta$ and $\bar\xi^\alpha$ (as well as
$\tilde\zeta_{\dot\alpha} \equiv \epsilon_{\dot\alpha\dot\beta}\tilde\zeta^{\dot\beta}$ and $\overline{\tilde\zeta}_{\dot\alpha}$) transform in the same way under Spin$(4).$
We can therefore construct spinorial derivatives
\begin{equation}
\label{spinorial:derivative:2}
D^\alpha
=
\frac{\partial}{\partial\xi_\alpha} + i\,{\tilde\zeta}_{\dot\beta}\,\tilde\sigma^{m\dot\beta\alpha}\,\partial_m,
\hskip 1cm
{\tilde D}_{\dot\alpha}
=
\frac{\partial}{\partial{\tilde\zeta}^{\dot\alpha}} + i\,\xi^\beta\,\sigma^{m}_{\beta\dot\alpha}\,\partial_m,
\end{equation}
and corresponding supercharges
\begin{equation}
\label{supercharges:2}
Q^\alpha
=
\frac{\partial}{\partial\xi_\alpha} - i\,{\tilde\zeta}_{\dot\beta}\,\tilde\sigma^{m\dot\beta\alpha}\,\partial_m,
\hskip 1cm
{\tilde Q}_{\dot\alpha}
=
\frac{\partial}{\partial{\tilde\zeta}^{\dot\alpha}} - i\,\xi^\beta\,\sigma^{m}_{\beta\dot\alpha}\,\partial_m,
\end{equation}
without invoking conjugated Grasmann variables. Then the set of equations:
\begin{equation}
\label{euclidian:basic:equation:1a}
D^\alpha\psi_{\alpha} + {\tilde D}_{\dot\alpha}\tilde\chi^{\dot\alpha} = 0,
\end{equation}
and
\begin{equation}
\label{euclidian:basic:equation:2a}
{\tilde D}^{ \dot\alpha}\psi_{\alpha} + D_{\alpha}\tilde\chi^{\dot\alpha} = 0,
\end{equation}
where $D_\alpha = D^{\beta}\epsilon_{\beta\alpha}$ and ${\tilde D}^{ \dot\alpha}= {{\tilde D}}{}_{\dot\beta}\epsilon^{\dot\beta\dot\alpha},$ imposed on ``analytic'', spinorial superfields
\begin{equation}
\psi_\alpha = \psi_\alpha(x,\xi,\tilde\zeta), \hskip 1cm  \tilde\chi^{\dot\alpha} = \tilde\chi^{\dot\alpha}(x,\xi,\tilde\zeta),
\end{equation}
is invariant with respect to both Spin$(4)$ and supersymmetric transformations (generated by (\ref{supercharges:2})) and implies the Dirac equation (\ref{eq:Dirac_equation_Euclidean}) on a
subspace satisfying the ``mass'' constraints
\begin{equation}
\label{mass:restricted}
\begin{split}
\frac14\left(\delta_\alpha^\beta \epsilon^{\dot\gamma\dot\alpha}\big[{\tilde D}_{\dot\alpha},{{\tilde D}}{}_{\dot\gamma}\big]
+
\epsilon_{\gamma\alpha}\big[D^\gamma,D^{\beta}\big]\right)\psi_\beta  & = m\psi_\alpha,
\\[4pt]
\frac{1}{4} \left( \delta^{\dot{\alpha}}_{\dot{\beta}} \epsilon_{\gamma\alpha}\big[ D^\alpha, D^\gamma\big]
+
\epsilon^{\dot\gamma\dot\alpha}\big[\tilde{D}_{\dot{\gamma}}, \tilde{D}_{\dot{\beta}}\big] \right)\tilde\chi^{\dot\beta}
& = m \tilde\chi^{\dot\alpha}.
\end{split}
\end{equation}

Nontrivial solutions of (\ref{euclidian:basic:equation:1a}), (\ref{euclidian:basic:equation:2a}) and (\ref{mass:restricted}) with $m = 0$ can be found (even if by ``brute force'', i.e.\ expanding
$\psi_\alpha$ and $\tilde\chi^{\dot\alpha}$ in a series of non-vanishing powers of $\xi$ and $\tilde\zeta$ and then working out and solving the resulting differential equations for the coefficient functions).
Notice that necessarily both $\psi_\alpha$ and $\tilde\chi^{\dot\alpha}$ are nonzero. Indeed, for $\tilde\chi^{\dot\alpha} = 0$ equations
(\ref{euclidian:basic:equation:1a}), (\ref{euclidian:basic:equation:2a}) imply
\begin{equation}
D^\alpha\psi_\alpha = 0, \hskip 1cm {\tilde D}_{\dot\beta}\psi_\alpha = 0,
\end{equation}
which is inconsistent since the anticommutator $\{D^\alpha,{\tilde D}_{\dot\beta}\}$ does not vanish.

\section{Almost-commutative geometry}
From a physicists point of view, the critical aspects of noncommutative geometry which make it so elegantly suited for the business of model building are the following:
\begin{enumerate}
\item  All physically relevant information pertaining to a manifold may be distilled within a short list of algebraic quantities, known as a spectral triple, $(\mathcal{A}, \mathcal{H}, \mathcal{D})$. Speaking informally, $\mathcal{A}$ is a (commutative) algebra of functions continuously defined on the manifold and which is faithfully represented as operators on a Hilbert space, $\mathcal{H}$, and $\mathcal{D}:\mathcal{H}\to \mathcal{H}$ is a Dirac operator.
\item Conversely, and under certain conditions, geometric information may be reconstructed from the data of an a priori given spectral triple. In particular, by allowing the algebra $\mathcal{A}$ to be noncommutative one recovers ``noncommutative" geometric information which is said to describe a ``noncommutative" manifold.
\item An action functional (and hence, Lagrangian) is then obtained by applying techniques of spectral theory (utilizing a heat kernal expansion) to the Dirac operator.  This stresses the importance of the role which the Dirac operator plays in this story, it essentially encodes the metric data of the model.

\end{enumerate}
A particularly interesting class of noncommutative geometries for physicists, due to their being a natural setting for the construction of gauge theories, are the so-called almost-commutative or (AC)-manifolds, a detailed description of which can be found in \cite{wvs}. Here, we find it sufficient to comment that such a manifold is actually described by the product of two spectral triples, the product being again a spectral triple. The first, loosely described above, and the second being a ``finite" spectral triple, $(\mathcal{A}_F, \mathcal{H}_F, D_F)$ consisting of a finite dimensional algebra, $\mathcal{A}_F$, represented on a finite dimensional Hilbert space, $\mathcal{H}_F$, and a symmetric matrix operator $D_F$, are combined into the ``total-space" spectral triple
\begin{equation}
(\mathcal{A}\otimes \mathcal{A}_F, \mathcal{H}\otimes \mathcal{H}_F, D_{AC}).
\end{equation}

With such an AC-geometry approach applied to our 4-dimensional Euclidean space, we have a total space Dirac operator of the form
\begin{equation}
\label{almost_commutative_Dirac_operator}
D_{AC} = {\cal D} \otimes \mathbf{1}_N + \gamma^5_E \otimes D_F,
\end{equation}
where ${\cal D}$ is the Euclidean Dirac operator defined in (\ref{dfn:Dirac_operator_Euclidean}), $\gamma^5_E$ is of the form given in (\ref{Euclidean_gamma_5}), and $D_F$ is a finite Dirac operator on $\mathbb{C}^N$, i.e. a Hermitian $N \times N$ matrix. Therefore, $D_{AC}$ can be explicitly written as a $4N \times 4N$ matrix, acting on bispinors of the form
\begin{equation}
\Psi = \left( \begin{matrix} \psi \\ \tilde{\chi} \end{matrix} \right),
\end{equation}
where
\begin{equation}
\psi = \left( \psi_{i \alpha} \right), \quad
\tilde{\chi} = \left( \tilde{\chi}^{\dot{\alpha}}_i \right), \quad
i = 1, \dots, N,
\end{equation}
and the Dirac equation can be written in the form
\begin{equation}
\label{eq:Dirac_equation_Euclidean_noncommutative}
\begin{split}
 i\,\tilde\sigma^{m\dot{\alpha} \alpha}\partial_m \psi_{i \alpha} + m \tilde{\chi}_i^{\dot{\alpha}} + \left( D_F \right)_{ij} \tilde{\chi}_j^{\dot{\alpha}} &= 0,
 \\[4pt]
i\,\sigma^m_{\alpha \dot{\alpha}}\partial \tilde{\chi}_i^{\dot{\alpha}} + m \psi_{i \alpha} - \left( D_F \right)_{ij} \psi_{j \alpha} &= 0.
\end{split}
\end{equation}
Consider now the algebra
\begin{equation}
\label{central:extended:D:algebra:a}
\begin{split}
\left\{D_i^\alpha,D_j^\beta\right\} &= 2\epsilon^{\alpha\beta}Z_{ij}, \hskip 2cm Z_{ij} = - Z_{ji},
\\[2pt]
\left\{\tilde D_{i\dot\alpha},\tilde D_{j\dot\beta}\right\} &= 2\epsilon_{\dot\alpha\dot\beta}\tilde Z_{ij}, \hskip 2cm \tilde Z_{ij} = - \tilde Z_{ji},
\end{split}
\end{equation}
together with
\begin{equation}
\label{central:extended:D:algebra:b}
\left\{D_i^\alpha,{\tilde D}{}^{\dot\alpha}_j\right\} =  2i\, \delta_{ij} \tilde\sigma^{m\dot\alpha\alpha}\,\partial_m,
\hskip 1cm
\left\{D_{i\alpha}, {\tilde D}_{j\dot\alpha}\right\} = 2i\,\delta_{ij}\sigma^{m}_{\alpha\dot\alpha}\,\partial_m,
\end{equation}
where $D_{i\alpha} = D_i^\beta\epsilon_{\beta\alpha}$ and ${\tilde D}^{\dot\alpha}_j = {\tilde D}_{j\dot\beta}\epsilon^{\dot\beta\dot\alpha}.$
It can be realized as an algebra of differential operators on a superspace with coordinates $(x^m,\xi_{i\alpha}, {\tilde\zeta}^{\dot\alpha}_i):$
\begin{equation}
\label{covariant:deriv:noncomm}
\begin{split}
D^\alpha_i &= \frac{\partial}{\partial\xi_{i\alpha}} + i {\tilde\zeta}^{i\dot\alpha} \tilde\sigma^{m\dot\alpha\alpha}\partial_m + Z_{ij}\xi^\alpha_j,
\\[2pt]
{\tilde D}_{i\dot\alpha} &= \frac{\partial}{\partial{\tilde\zeta}^{\dot\alpha}_j} + i \xi^{\alpha}_i\sigma^m_{\alpha\dot\alpha}\partial_m + {\tilde Z}_{ij}{\tilde\zeta}_{j\dot\alpha}.
\end{split}
\end{equation}
The corresponding supercharges, anticommuting with derivatives (\ref{covariant:deriv:noncomm}), have the form:
\begin{equation}
\label{supercharges:noncomm}
\begin{split}
Q^\alpha_i &= \frac{\partial}{\partial\xi_{i\alpha}} - i {\tilde\zeta}^{i\dot\alpha} \tilde\sigma^{m\dot\alpha\alpha}\partial_m - Z_{ij}\xi^\alpha_j,
\\[2pt]
{\tilde Q}_{i\dot\alpha} &= \frac{\partial}{\partial{\tilde\zeta}^{\dot\alpha}_j} - i \xi^{\alpha}_i\sigma^m_{\alpha\dot\alpha}\partial_m - {\tilde Z}_{ij}{\tilde\zeta}_{j\dot\alpha}.
\end{split}
\end{equation}

If we postulate equations of the form
\begin{equation}
\label{eq:extended_supersymmetry}
\begin{split}
D_i^\alpha \psi_{j \alpha} + \tilde{D}_{j \dot{\alpha}} \tilde{\chi}_i^{\dot{\alpha}} &= 0,
\\[2pt]
\tilde{D}_i^{\dot{\beta}} \psi_{i \alpha} + D_{i \alpha} \tilde{\chi}_i^{\dot{\beta}} &= 0, 
\end{split}
\end{equation}
then, using (\ref{central:extended:D:algebra:b}), we can conclude that solutions of (\ref{eq:extended_supersymmetry}) satisfy the Dirac equation, (\ref{eq:Dirac_equation_Euclidean_noncommutative}),
provided that the ``mass" conditions
\begin{equation}
\label{noncom:mass:1}
\begin{split}
\left(m\delta_{ij} + (D_F)_{ij}\right)\tilde\chi^{\dot\alpha}_j
&=
\frac12\left(\delta^{\dot\alpha}_{\dot\beta} D^\alpha_i D_{j\alpha} + \delta_{ij} {\tilde D}^{\dot\alpha}_k {\tilde D}_{k\dot\beta}\right)\tilde\chi^{\dot\beta}_j,
\\[2pt]
\left(m\delta_{ij} - (D_F)_{ij}\right)\psi_{j\alpha}
&=
\frac12\left(\delta^\beta_\alpha {\tilde D}_{i\dot\alpha}{\tilde D}^{\dot\alpha}_j + \delta_{ij}D_{k\alpha}D^\beta_k\right) \psi_{j\beta},
\end{split}
\end{equation}
are satisfied. With the help of (\ref{central:extended:D:algebra:a}) equation (\ref{noncom:mass:1}) can be alternatively presented as
\begin{equation}
\label{noncom:mass:2a}
\left(m\delta_{ij} + (D_F)_{ij}- Z_{ij}\right)\tilde\chi^{\dot\alpha}_j
=
\frac14\left(\delta^{\dot\alpha}_{\dot\beta} \epsilon_{\beta\alpha}\big[D^\alpha_i, D^\beta_j\big]
+
\delta_{ij} \epsilon^{\dot\gamma\dot\alpha}\big[{\tilde D}_{k\dot\gamma},{\tilde D}_{k\dot\beta}\big]\right)\tilde\chi^{\dot\beta}_j,
\end{equation}
and
\begin{equation}
\label{noncom:mass:2b}
\left(m\delta_{ij} - (D_F)_{ij} -\tilde Z_{ij}\right)\psi_{j\alpha}
=
\frac14\left(\delta^\beta_\alpha \epsilon^{\dot\beta\dot\alpha}\big[{\tilde D}_{i\dot\alpha},{\tilde D}_{j\dot\beta}\big]
+
\delta_{ij}\epsilon_{\gamma\alpha}\big[D_{k}^\gamma,D^\beta_k\big]\right) \psi_{j\beta}.
\end{equation}
The simplest solutions of these equations (and, most likely, the only consistent with (\ref{eq:extended_supersymmetry}), although the general proof of this claim is still missing)
correspond to a situation in which both the l.h.s.\ and the r.h.s.\ of (\ref{noncom:mass:2a}) and (\ref{noncom:mass:2b}) vanish.
This implies that the constructed framework, which reconciles non-commutative geometry with supersymmetry in a simple setting, requires the finite
part of the Dirac operator (\ref{almost_commutative_Dirac_operator}) to be antisymmetric and expressible through central charges of the algebra
(\ref{central:extended:D:algebra:a},\ref{central:extended:D:algebra:b}) as
\begin{equation}
 (D_F)_{ij} = Z_{ij} = - \tilde Z_{ij}.
\end{equation}

It is worth mentioning, that in the usual development via the AC-geometry approach to noncommutative geometry, the finite spectral triple only contains data pertaining to the fermionic particle content of the model.  The bosonic content of the theory, or gauge fields, are then given by the inner fluctuations which arise through consideration of Morita equivalences of the algebra.  The Morita (self-)equivalent total-space spectral triple is then comprised of the algebra, Hilbert space, and the ``fluctuated" Dirac operator taking into account the gauge fields.
While we have seen that gauge fields arise naturally through the ``factorization" procedure which we have herein described, one could also consider the implications of factorizing the fluctuated Dirac operator.  This possibility is almost certainly worthy of further investigation.

\end{document}